\documentclass[runningheads]{llncs}
\usepackage{graphicx}
\usepackage{sgame}
\usepackage{booktabs}
\usepackage{amsmath}
\usepackage{amssymb}

\usepackage{biblatex} 
\begin{document}
\title{Evaluation of an Anomaly Detector for Routers using Parameterizable Malware in an IoT Ecosystem
\thanks{The work was funded in part by Spiros Mancoridis’ Auerbach Berger Chair in Cybersecurity.}}
\author{John Carter\inst{1}\orcidID{0000-0003-4391-0269} \and
Spiros Mancoridis\inst{1}\orcidID{0000-0001-6354-4281}}
\authorrunning{J. Carter et al.}
\titlerunning{Evaluation of an Anomaly Detector using Parameterizable Malware}
\institute{Drexel University, Philadelphia PA 19104, USA \\
\email{jmc683@drexel.edu}
}
\maketitle             

\begin{abstract}
This work explores the evaluation of a machine learning anomaly detector using custom-made parameterizable malware in an Internet of Things (IoT) Ecosystem. It is assumed that the malware has infected, and resides on, the Linux router that serves other devices on the network, as depicted in Figure \ref{fig:ecosystem}. This IoT Ecosystem was developed as a testbed to evaluate the efficacy of a behavior-based anomaly detector. The malware consists of three types of custom-made malware: ransomware, cryptominer, and keylogger, which all have exfiltration capabilities to the network. The parameterization of the malware gives the malware samples multiple degrees of freedom, specifically relating to the rate and size of data exfiltration. The anomaly detector uses feature sets crafted from system calls and network traffic, and uses a Support Vector Machine (SVM) for behavioral-based anomaly detection. The custom-made malware is used to evaluate the situations where the SVM is effective, as well as the situations where it is not effective.

\end{abstract}

\section{Introduction}
Malware detection on small, resource-constrained devices has emerged as an important area of research, as IoT devices have grown in popularity. Since these devices have limited resources \cite{10.1145/3139937.3139944}, malware detection software running on them must be efficient and lightweight, yet accurate and useful. This work focuses on creating and deploying custom-made parameterizable malware on a router in an IoT ecosystem to evaluate an anomaly detector's effectiveness in detecting the presence of malware. The parameterization of the malware enables the conditions on the router to vary, which provides a variety of data with which to train and test the anomaly detector. We show that while the SVM is incredibly effective and practical as an anomaly detector on IoT devices due to its low resource consumption and high accuracy, the parameters of the malware can be adjusted to decrease the effectiveness of the SVM.

\begin{figure}
    \centering
    \includegraphics[width=\columnwidth]{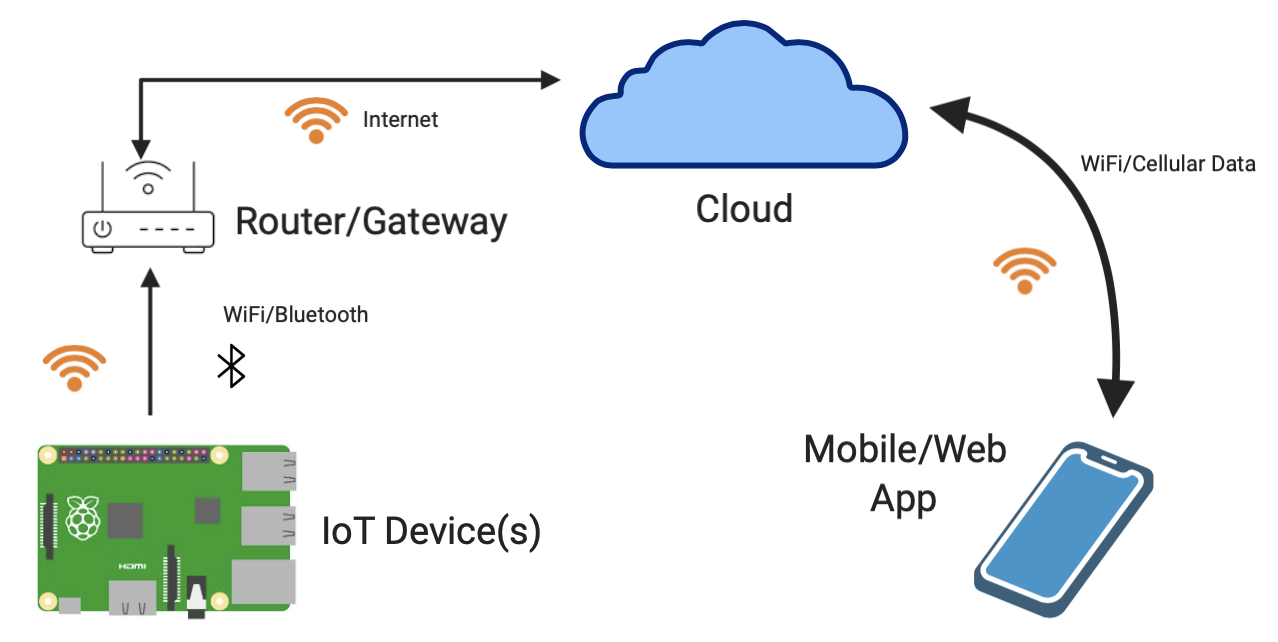}
    \caption{The IoT Ecosystem\label{fig:ecosystem}}
\end{figure}

\indent Routers can be described as IoT devices that do not require constant human oversight, and have limited behavioral patterns and resources. Their limited behavior make anomaly detection simpler, but the anomaly detection we describe in this work could likely be ported to non-IoT devices as well. Since routers are used to connect devices on a local area network (LAN) to the Internet, their security is critical. If a router is infected with malware, there is a high probability that the malware can spread quickly to other devices on the network. This idea underpins one of the important reasons to work on the topic of securing routers. If it is possible to make a router resilient to malware, then it is possible that the router could act as a firewall to prevent malware from infecting devices connected to the router's network. In this work, the router we are working with is a Raspberry Pi 3, called Pi-Router, that has been configured to work as a router using the \textit{hostapd} package. This package allows the Raspberry Pi to work as a wireless access point. The Raspberry Pi gives a realistic picture of working on an IoT device due to its own limited resources, and the fact that it runs on a Linux distribution. This provides a useful environment in which to develop lightweight malware detection systems that can be ported to similar Linux-based IoT devices.

\indent In order to have a diverse set of fully-functional and parameterizable malware samples, we have created three types of malware for this work: ransomware, cryptominer, and keylogger, all with remote exfiltration capabilities. Some of these malware samples, such as the keylogger, may not be commonly found on routers, but including it in this research demonstrates the breadth of different malware examples that could infect any IoT device. Often, malware caught "in the wild" fails to run well, or at all, for a variety of reasons. These can include outdated code, attempts to connect to a server that no longer exists, and other reasons. This situation can make malware samples from the wild less useful to train and test a malware detector, which creates a need for custom-made malware that emulates how different malware families behave.

\begin{figure}
    \centering
    \includegraphics[width=\columnwidth]{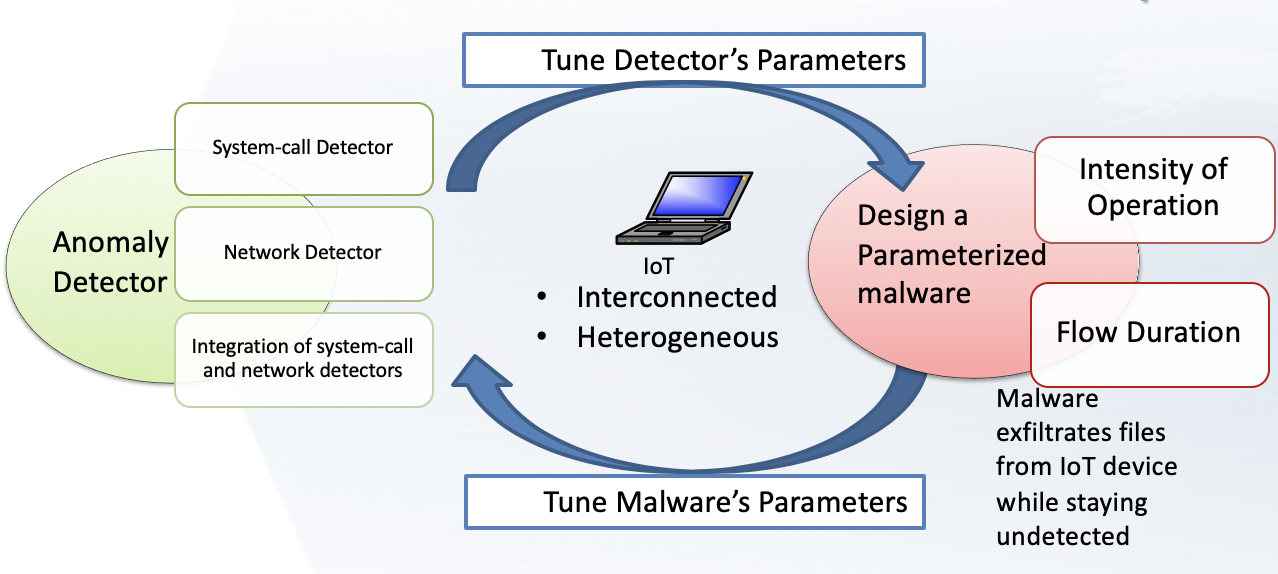}
    \caption{Parallel Development of an Anomaly Detector and Malware Samples\label{fig:overview}}
\end{figure}

\indent The malware created all have parameterizable exfiltration rates, which means they can be tuned to attempt to elude the anomaly detector. These degrees of freedom on the malware samples are essential to emulate the adversarial relationship between the malware and the malware detection software. The parameterization of the malware is depicted on the right side of Figure \ref{fig:overview}. The specific functionality and capabilities of each of the malware samples will be discussed further in Section 3.

\indent A SVM is used for router behavioral anomaly detection. The SVM was chosen because it can be trained effectively with less data than other machine learning models, such as neural networks, and it is more efficient to run on a resource-constrained IoT device. The detector was trained on three types of data. The rest of the paper demonstrates how it is possible to detect malware running on the Pi-Router by first training the anomaly detector on kernel-level system call data, training the detector on network traffic data, and lastly, training the detector on a combination of the two, which is depicted on the left side of Figure \ref{fig:overview} \cite{MALWARE19}. Each of these anomaly detectors are accurate in classifying data, but we will show that the detector trained on the combination of data generally performs the best.

\section{Related Work}
This work draws on prior research in the areas of anomaly detection and malware detection on IoT devices. Previous work focused on specific IoT devices with predictable behavior, such as Amazon's Alexa running on a Raspberry Pi. This work approaches the problem more broadly in the context of routers infected with custom-made parameterizable malware \cite{MALWARE19} \cite{Noorani2019AutomaticMD}.

\indent It has been shown that sequences of system calls and network traffic data provide useful and explanatory feature sets for classifiers, as they provide insight to the programs running on a machine and can often be used to differentiate between a period of benign behavior, and a period of malware infection \cite{MALWARE19}. 

\indent System call sequences have been shown to be effective as a feature set in many architectures due to the fact that they are one of the best indicators of what is happening on a machine at runtime \cite{MALWARE19} \cite{8659366}. 
Each process running on a machine uses system calls to request resources from the OS kernel, which means the programs running normally are likely to make similar system calls each time they run. 
Because of this, it will be more obvious when a new program starts running because it will likely either issue a different set of system calls or issue the same system calls in a different order. 
It has been suggested, by results in prior work, that an anomaly detector based solely on system call traces performs well enough on its own to be effective in malware detection \cite{MALWARE19}.

\indent Network traffic data have also been shown to be effective as a feature set for behavior anomaly detection \cite{Doshi_2018} \cite{8093473}. 
This type of network traffic analysis is also the basis of network intrusion detection systems (NIDS)~\cite{DBLP:journals/corr/abs-1901-03407,6524462,nids1}.
Network traffic provides insight into all the communication a device is having on the network, including where it is sending data and from where it is receiving data \cite{MALWARE19}.

\indent This research is intended to build on the prior work discussed, and broaden its application to a more general IoT platform. Another key distinction of this work is the design and use of custom-made parameterizable malware to aid in the evaluation of behavioral-based anomaly detectors.

\section{Malware}
Three types of malware are used in the experimentation, which together provide a variety in the breadth of the malware. Some of the malware are more computationally expensive on the device and issue many system calls, while others issue fewer system calls but exfiltrate more information at varying speeds, which contributes to higher overall network traffic. All of the malware used in this research is custom-made, which provides more freedom to create interesting types of malware that are diverse in their execution behavior, yet stay true to behaviors that would be exemplified in real malware samples. In addition, the malware used in this research each have varying degrees of freedom, that include the rate of data exfiltration, as well as the size of the data exfiltration. These degrees of freedom are important because they enable the malware to adapt to changing conditions on the host prompted by malware detection software, as well as provide a better basis to show situations where the malware detection is successful, and situations where it is less successful.

\subsection{Keylogger} 
The first type of malware used in this research is a keylogger, which tracks all of the key presses on the device, and saves them to a buffer. The key presses can then be exfiltrated to another machine. The speed and the size of the exfiltration are set by user-defined parameters, which means an adversary can adjust the malware to attempt to avoid detection by an anomaly detector. The speed of the exfiltration refers to the rate of exfiltration, which can be defined as the interval at which the buffer of key presses, or a subset of them, are removed from the buffer and sent to a remote machine. The size of the exfiltration, in this case, refers to the number of key presses to send with each exfiltration packet to a remote machine. This allows the adversary to adapt the malware in order to evade detection and possible mitigation strategies implemented by the malware detection. Although a keylogger is not usually deployed on a router, this malware family was included in this work to provide a larger breadth of malware behaviors with which to train a more effective IoT behavioral anomaly detector.

\subsection{Ransomware}
The ransomware malware uses the Python cryptography library to encrypt a file system, and exfiltrate the contents to a remote host on the network. The exfiltration happens during the encryption process. An encryption key is created, and then the malware traverses the file system. For each file it finds, it first sends a copy of it to the remote host using secure copy (scp), and then encrypts it and continues to the next file. The decryption functionality is also provided. The location to save the file on the remote host is user-specified. The user can also provide an exfiltration interval, which will insert a delay in between exlfiltrating individual files. Similar to the rate of exfiltration on the keylogger, this allows an adversary to become more inconspicuous, since copying and sending files at a slower rate will likely draw less attention to the malware than performing the same process rapidly. 

\subsection{Cryptominer}
Lastly, the cryptominer malware is a simple coin mining script, that runs a mining simulation to emulate the computational cost of a real cryptominer. The user specifies a remote host to send the new hash that was mined on the host after the completion of the mining process. Similar to the ransomware, the user can specify an exfiltration interval, which will insert a delay between the mining process and the exfiltration of the calculated hash. While the behavior of the two previous malware discussed is often easily traceable by both system call data and network traffic data, the cryptominer malware is more easily traceable by system call data, due to the heavy computation cost of the mining process, and the relatively few packets being sent over the network as a result of its execution.

\section{Anomaly Detection Model}
The feature sets for the anomaly detection model are extracted from sequences of system calls and network traffic flows on the Pi-Router. This includes any system calls executed on the Pi-Router during data collection, as well as any packets sent to or from the Pi-Router during its execution.

\subsection{System Calls}
\indent System calls indicate the activity of each running process on the machine. Therefore, when a malware sample starts to execute, its process will likely make different system calls than previously seen and will be useful in detecting an anomaly present on the machine.

\indent The pre-processing step for system calls is similar to \cite{8659366}, where a sequence of system calls collected in a window size of length \emph{L} is treated as an observation. Then, a bag-of-\emph{n}-grams approach \cite{7272922}, \cite{8323956}, \cite{924295} is used to group the system calls and create the feature vector $\textbf{x} \in \mathbb{R}^p$. This can be described as the number of times a system call \emph{n}-gram sequence was observed in an observation window of length \emph{L} \cite{MALWARE19}. An \emph{n}-gram length of $\emph{n}=2$ was used when the results of the classifier were compiled, although this is another parameter of the data processing code that can be changed by the user. Other values of \emph{n}, such as $\emph{n}=3$, were used in the experimentation phase as well, but did not yield significantly better results.

\subsection{Network Traffic}
Network traffic features are extracted from network flows collected by CICFlowMeter. CICFlowMeter is a package in the Python Package Index (PyPi) that listens to network traffic on a device, generates bidirectional network flows, and then extracts features from these flows. In the network traffic data collected, one packet sent or received by the Pi-Router counts as one observation. 

\indent One issue that arises when using network traffic as a feature set is that many of the packets that are sent by normal non-malicious applications on the Pi-Router are also included in the malware dataset. This results in these benign packets being labeled as malicious data, which yields an inseparable dataset \cite{MALWARE19}. This issue is resolved by grouping the network traffic data into \emph{m}-second intervals, which results in more finely-grained malware traces that distinguish them from the benign traces \cite{MALWARE19}. The key idea here is to find the optimal value of \emph{m} so that the bin is large enough to collect enough packets to determine their source or destination program, but small enough so that packets from other programs are excluded, making each packet sequence more distinct. As with the bag-of-\emph{n}-grams approach used in the system call processing, the value of \emph{m} is a tunable parameter that can be adjusted by the user. During the experimentation phase of this work, a few different values of \emph{m} were used, and it was found that smaller values of \emph{m} are better for classifying our malware samples.

\subsection{Principal Component Analysis}
\begin{figure}
    \centering
    \includegraphics[width=\columnwidth]{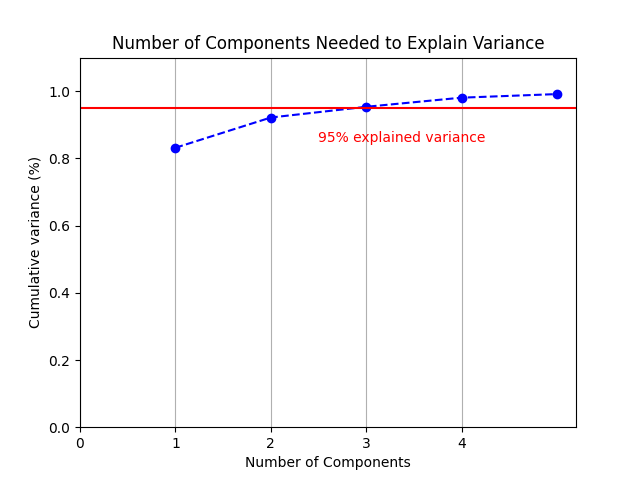}
    \caption{Explained Variance for the Cryptominer Combined Feature Set\label{fig:pca}}
\end{figure}

After the pre-processing mentioned above was applied to the datasets, approximately 2600 features were extracted from the system call dataset, and 237 features were extracted from the network traffic dataset. Combining these features brings the total number of features in the feature set to approximately 2800 features. Principal Component Analysis (PCA) was then used to reduce the dimensionality of the created feature space, similar to work by \cite{MALWARE19}. The number of components to use was determined by calculating the explained variance of the components, and selecting the number of components that explain at least 95\% of the variance in the dataset. Figure \ref{fig:pca} depicts the number of components needed to explain 95\% of the variance for the Cryptominer combined feature set, which in this specific case is four components.

The anomaly detection model uses a  Support Vector Machine (SVM), which finds a hyperplane that separates the training data from origin with the largest-possible margin. The objective function is thus maximizing the margin, or the distance from the origin to the hyperplane. 

\section{Experimental Setup}
\subsection{Pi-Router}
\begin{figure}
    \centering
    \includegraphics[width=\columnwidth]{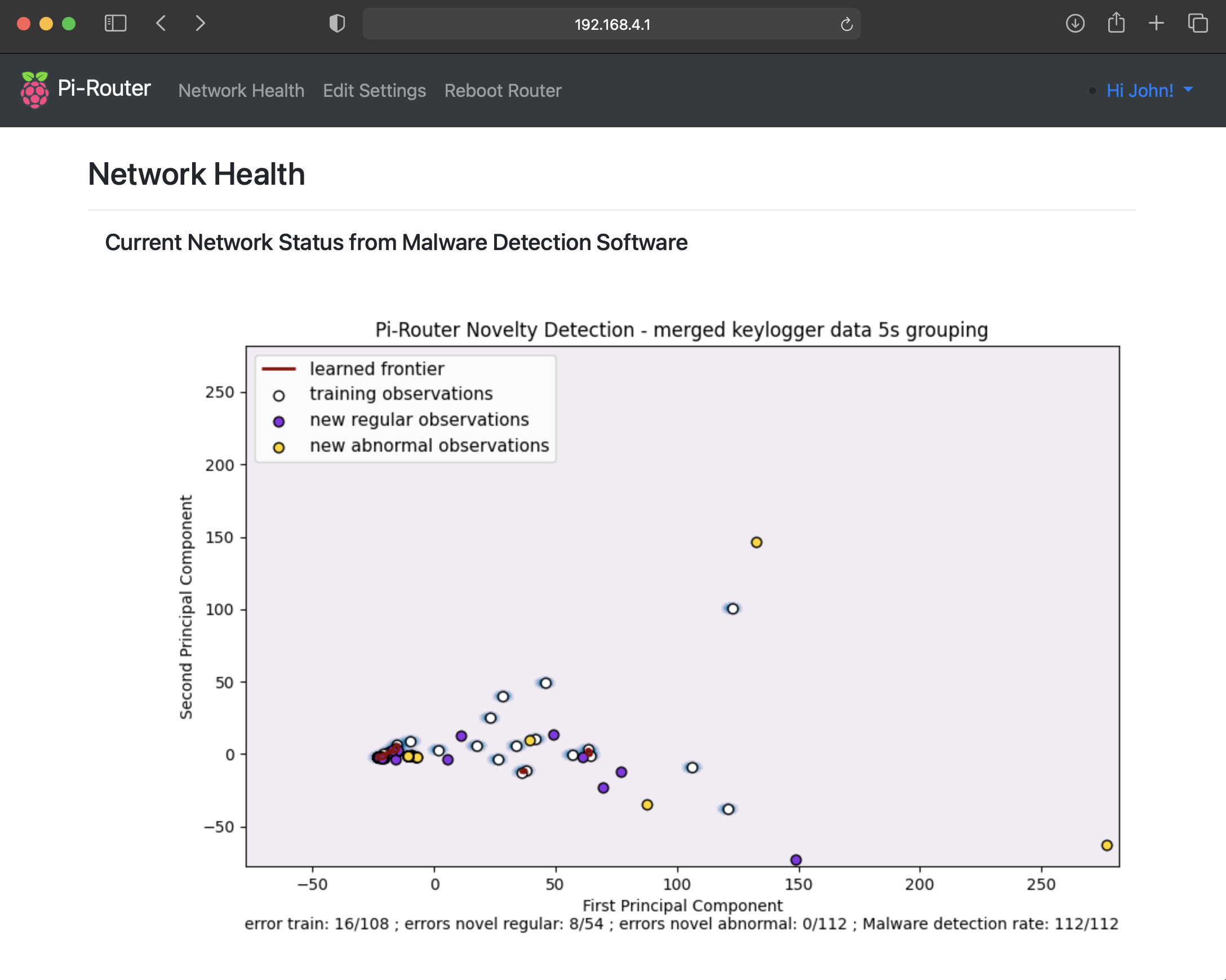}
    \caption{Example Screenshot of the Pi-Router UI\label{fig:webserver}}
\end{figure}

Most of the setup for the experiments conducted is focused on the Pi-Router. The Pi-Router is a Raspberry Pi 3 running Raspberry Pi OS (formerly Raspbian), which is based on the Debian Linux distribution. This Raspberry Pi has been configured to act as a wireless access point by using the software package \textit{hostapd}, which is a user-space daemon that allows the network interface card (NIC) to act as an access point. A Flask web server running as a Linux service was created on the Pi-Router to act as a portal for users to configure the router and track the health of the network by running the malware detection software and showing the results. Figure \ref{fig:webserver} shows how the Network Health section might look to the user on the Pi-Router web server, where the graphs show the results of running the SVM on current system call and network traffic data. 

\indent The web server presents the essence of the usability of this system, in that it allows users to monitor the behavior and overall health of their network and track anomalous behaviors caused by malware.

\subsection{Pi-Network and Connected Devices}
The Pi-Router's network is the Pi-Network, which has two other hosts connected to it currently: a Pi-Camera and the attacker machine. The Pi-Camera in this work is simply another host on the network that generates traffic unrelated to the router. The attacker is a laptop running Ubuntu 20.04 and is responsible for receiving all of the exfiltrated data sent by the malware. This includes key presses sent from the keylogger, files copied by the ransomware, and hashes sent by the cryptominer. As a result, the relevant network traffic would often originate from the Pi-Router or the attacker, and be received by the other. Most of the exfiltration communication was one-way, from the Pi-Router to the attacker.

\section{Experimental Results}
\begin{figure}
    \centering
    \includegraphics[width=\columnwidth]{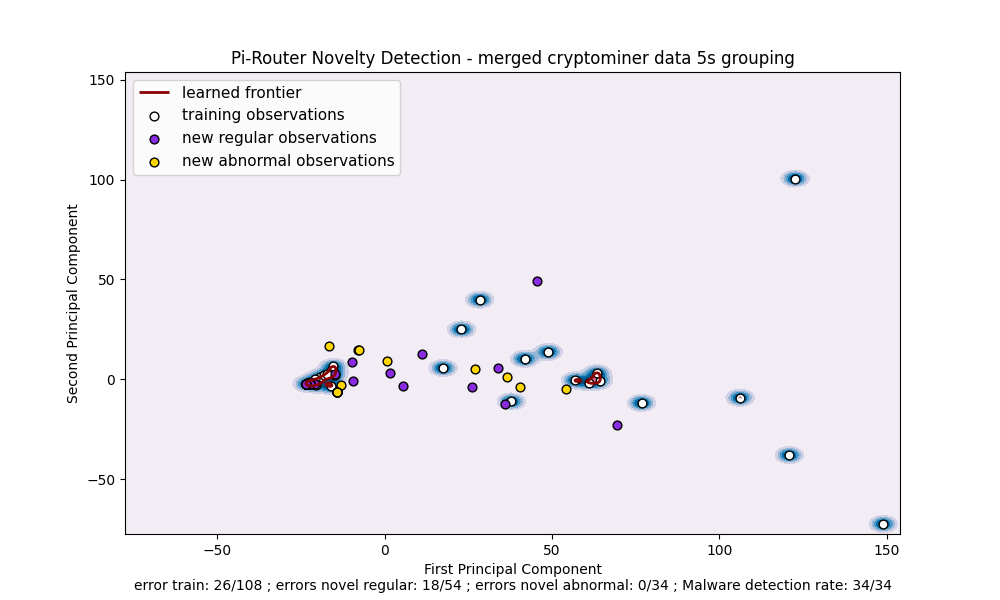}
    \caption{Visualization of first two Principal Components on Cryptominer data using a 5s window size\label{fig:keylogger_classification}}
\end{figure}

The classification experiments were conducted with the three types of malware, each with different window sizes, or values of \emph{L}, used in the SVM. Each of these was carried out for the three types of features: system calls, network traffic, and a combination of the two. In Figure \ref{fig:keylogger_classification}, we show a visualization of the classification process using the SVM, in which the first two principal components derived from PCA are shown. In this example, the SVM is classifying keylogger data using a five-second window size. The goal is to have as many new abnormal observations outside of the learned frontier as possible, because they will be easier to identify and isolate from the benign data and thus be better indicators of a possible malware infection.

\begin{figure}[!htbp]
  \centering
  \begin{tabular}{c c c}
    \includegraphics[width=0.3\textwidth]{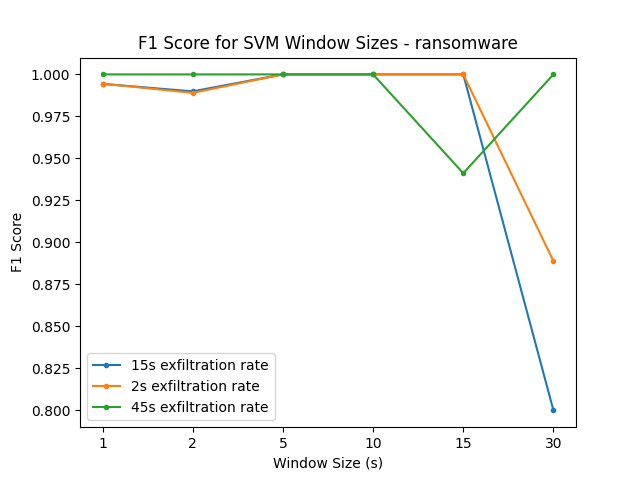} &
    \includegraphics[width=0.3\textwidth]{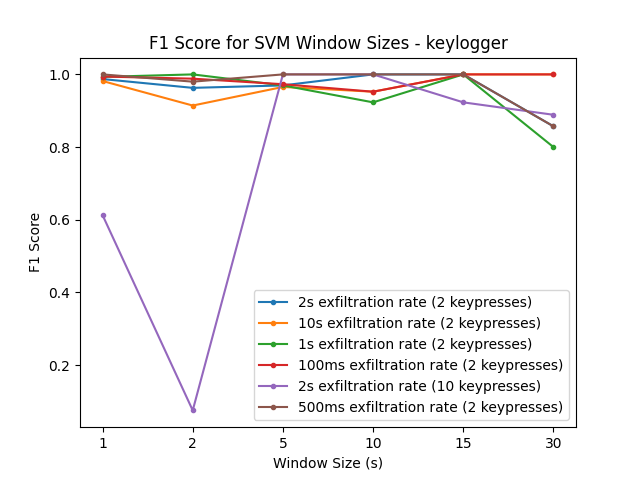} &
    \includegraphics[width=0.3\textwidth]{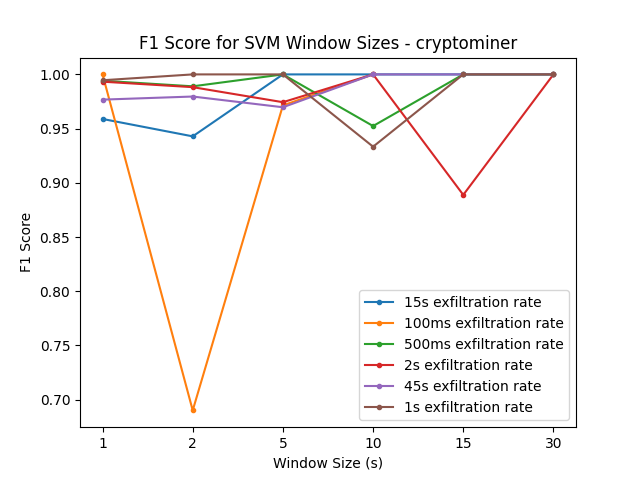}
  \end{tabular}
  \caption{A comparison of F1 Scores using different Window Size values (values of \emph{L}) 
  \label{fig:f1}}
\end{figure}

\indent After some experimentation, it was found that a $\emph{L}=5$ seconds window size was optimal for classification accuracy. Other window sizes yielded suboptimal results for each feature set. Figure \ref{fig:f1} shows the comparison between F1 score and window size for the three types of malware, in which we see that for all three types of malware, a $\emph{L}=5$ seconds window size had the best classification performance of the 6 different window sizes.

\begin{figure}[!htbp]
  \centering
  \begin{tabular}{c c c}
    \includegraphics[width=0.3\textwidth]{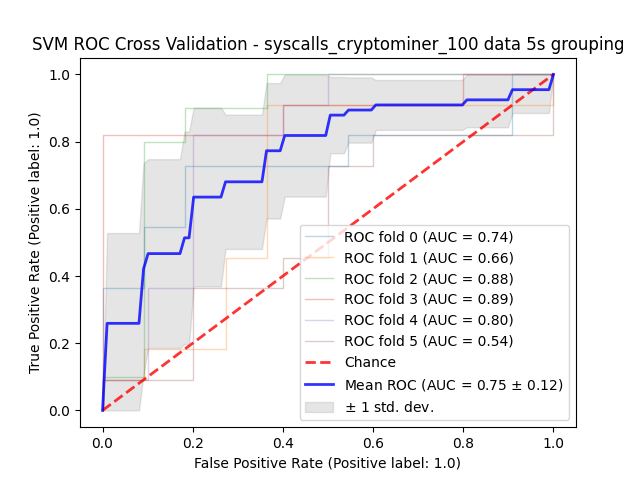} &
    \includegraphics[width=0.3\textwidth]{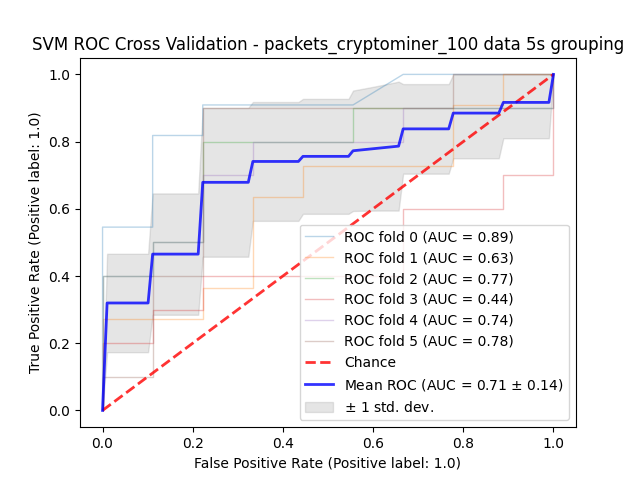} &
    \includegraphics[width=0.3\textwidth]{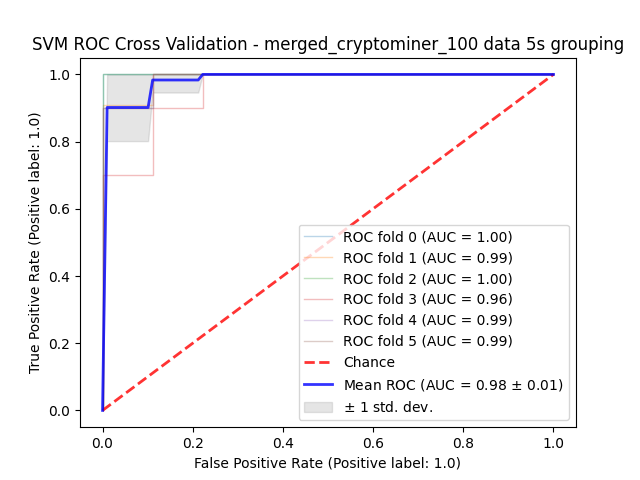}
  \end{tabular}
  \caption{As with the other types of malware, the combined feature set outperforms the feature sets comprised of system call data and network traffic data individually. Here, we demonstrate this using the Cryptominer malware with a Window Size of 5s and an exfiltration rate of 100ms. The first figure uses system call features, the second uses network traffic features, and the third uses both types of features.
  \label{fig:roc_comparison}}
\end{figure}

\indent In addition, we found that using a classifier trained on both the system call feature set and network traffic feature set generally outperformed classifiers trained on each type of data individually. While system call data can provide a lot of information relating to the behavior of a process, the network traffic data often augments that behavior data and makes the classification results more conclusive. Figure \ref{fig:roc_comparison} shows an example of the merged data providing better features than system call and network traffic data individually using cryptominer data with a five second window size. In this case, both the classifier trained on system call data and the classifier trained on network traffic data each only had a mean AUC value of 0.75 or less, while the combined data had a mean AUC value of 0.98, which is a significant improvement.

In all three types of malware, the combined dataset of system calls and network traffic generally outperformed the system call and network traffic classifiers individually. 

\begin{figure}[!htbp]
  \centering
  \begin{tabular}{c c c}
    \includegraphics[width=0.3\textwidth]{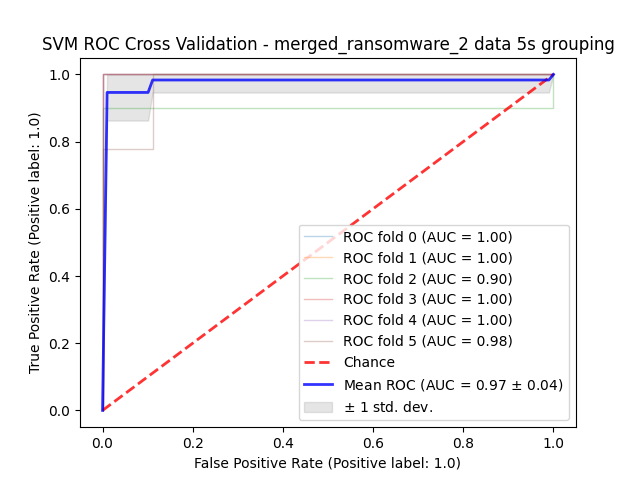} &
    \includegraphics[width=0.3\textwidth]{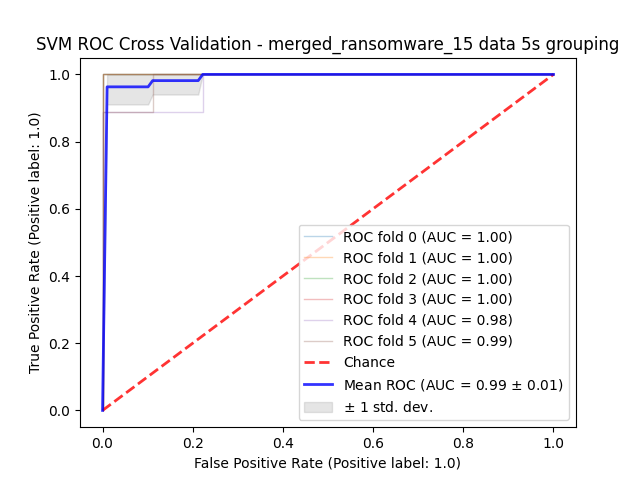} &
    \includegraphics[width=0.3\textwidth]{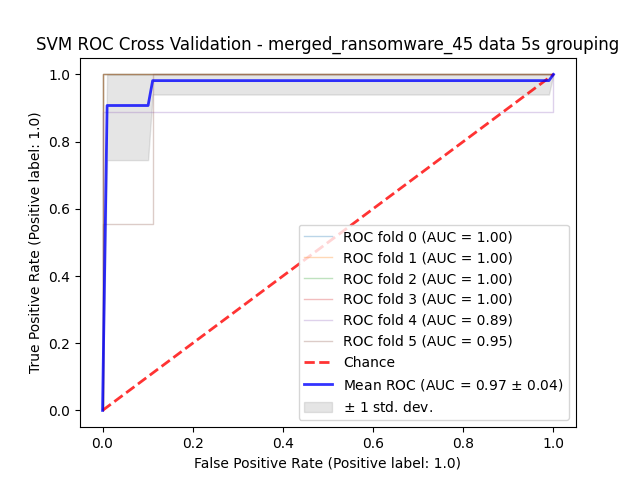}
  \end{tabular}
  \caption{ROC Curves for Ransomware with a window size of 5 seconds and a 2s exfiltration rate, a 15s exfiltration rate, and a 45s exfiltration rate. 
  \label{fig:roc_ransomware}}
\end{figure}

\begin{figure}[!htbp]
  \centering
  \begin{tabular}{c c c}
    \includegraphics[width=0.3\textwidth]{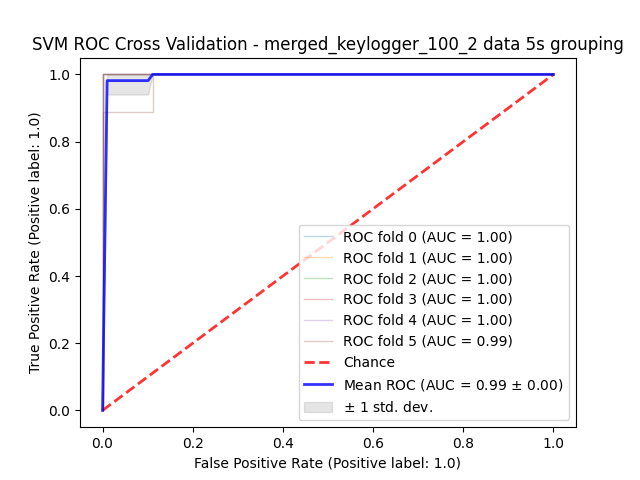} &
    \includegraphics[width=0.3\textwidth]{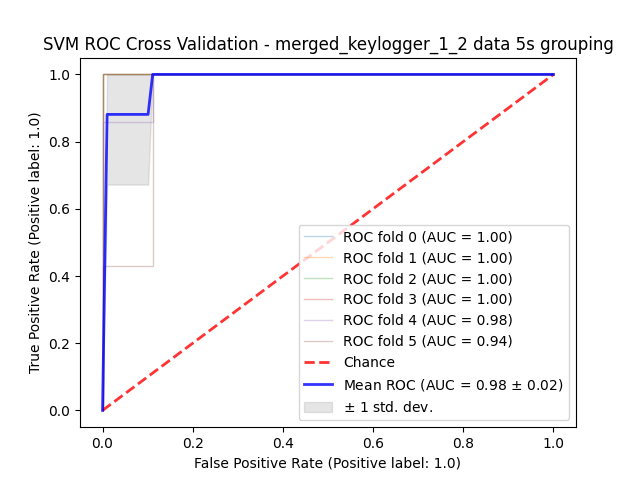} &
    \includegraphics[width=0.3\textwidth]{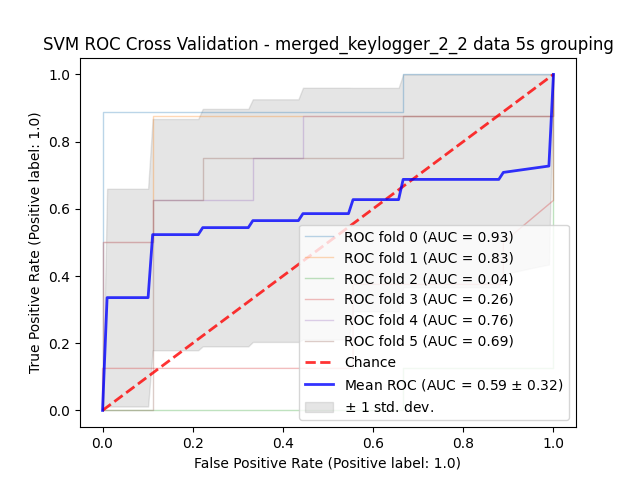}
  \end{tabular}
  \caption{ROC Curves for Keylogger with a window size of 5 seconds and a 100ms exfiltration rate, a 1s exfiltration rate, and a 2s exfiltration rate. 
  \label{fig:roc_keylogger}}
\end{figure}

\begin{figure}[!htbp]
  \centering
  \begin{tabular}{c c c}
    \includegraphics[width=0.3\textwidth]{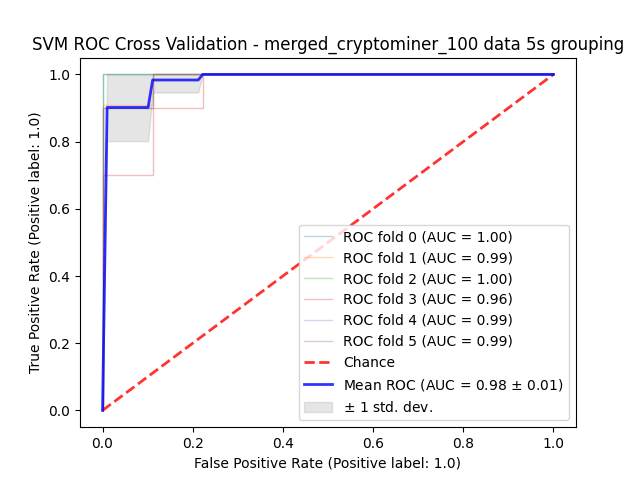} &
    \includegraphics[width=0.3\textwidth]{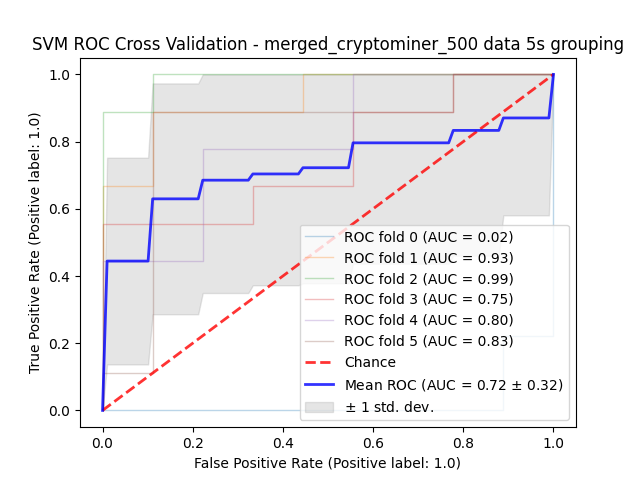} &
    \includegraphics[width=0.3\textwidth]{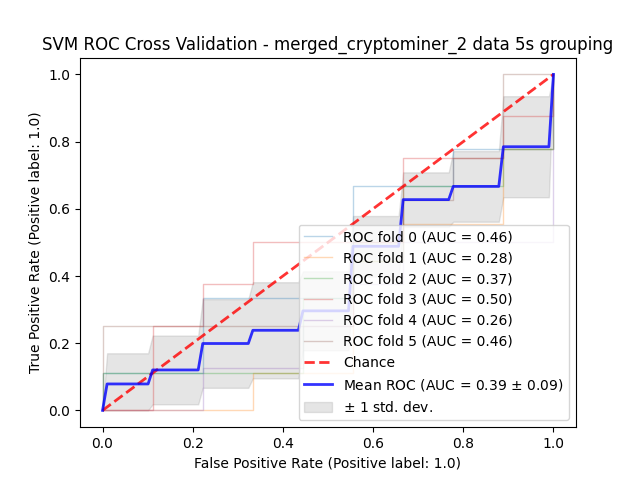}
  \end{tabular}
  \caption{ROC Curves for Cryptominer with a window size of 5 seconds and a 100ms exfiltration rate, a 500ms exfiltration rate, and a 2s exfiltration rate. 
  \label{fig:roc_cryptominer}}
\end{figure}

\indent The parameters provided in the malware samples greatly affect the classification effectiveness of the SVM. Specifically, the rate of exfiltration parameter in each malware sample is instrumental in its detection. In general, as one might expect, the faster the exfiltration rate, the easier the detection process becomes. Rapidly issuing system calls and sending packets results in more conspicuous malware, while slower malware that is idle in between actions for longer periods of time is much harder to detect. Figures \ref{fig:roc_ransomware}-\ref{fig:roc_cryptominer} demonstrate this idea using ROC-AUC results for each malware with different parameter settings, again all with a window size of 5 seconds. Here, the results indicate that the keylogger exfiltration rate did play an important part in the classification results, but the number of system calls it made and the number of packets it sent still made the malware easy to detect. In contrast, the exfiltration rates of the other two malware were useful in their ability to elude the malware detection, since as the malware continued to run with slower exfiltration rates they were much harder to detect.

\indent The differences in results shown in Figures \ref{fig:roc_ransomware}-\ref{fig:roc_cryptominer} demonstrate the impact the degrees of freedom have on the detection of the malware. By changing one tunable parameter on the malware, they can be either easily detectable with a high AUC value, or hardly detectable, with an AUC value similar to chance classification. By using these custom-made malware samples, we can provide a larger set of data with more variations to show where the malware detection excels, and where it fails, to improve the overall classification process.

\indent The co-design of parameterizable malware and anomaly detectors is useful to design a robust detector that can detect elusive, inconspicuous, malware.

\section{Conclusion \& Future Work}
This research focuses on creating parameterizable malware and building an anomaly detector to detect the malware using machine learning. The malware lives on a Linux router in an IoT Ecosystem. The ecosystem was designed and created to be a testbed to enable evaluation of anomaly detectors using custom-made malware. The malware is useful because it allows for more varied data to be generated while still emulating its respective malware family. This method of malware detection is useful for IoT devices because it does not rely on prior knowledge of the environment of the device, and is also very resource efficient, which is essential if the malware detection is going to run on the IoT device itself.

\indent In this research, we created a real IoT Ecosystem, which in this work focuses primarily on the Pi-Router. The Pi-Router is a fully-functional wireless access point that provides a network that devices can connect to locally. We also created malware that is fully-functional to run on the Pi-Router. The malware provide several degrees of freedom with which to create varied data and evaluate the anomaly detection model under different environment conditions. Lastly, we demonstrated the creation of a behavioral anomaly detection system, which is designed to detect malicious software running on the Pi-Router and on IoT devices in general. The anomaly detection system was trained on three types of data: system call sequences, network traffic data, and a combined dataset comprised of system calls and network traffic data. Our results indicate that a classifier trained on the combined data with a window size of \emph{L}=5 seconds generally outperformed the other classifiers that were used. We also found that the ability of the malware samples to elude the anomaly detector was generally dependent on the exfiltration rate of the malware.

\indent We found that this classifier is very useful for malware detection on IoT devices. In addition, the custom-made malware is useful for identifying situations where the SVM was successful and situations where it was not. The degrees of freedom of the malware were significant in providing a way to quickly generate different data from the same malware family to test the classifier's ability to adapt to changing malware behavior. We plan to continue and expand on this research by using the data obtained from the IoT Ecosystem to train a generator and a discriminator in a Generative Adversarial Network (GAN) \cite{nids1}. Feeding the real data to the GAN initially will save training time, at which point the generator will have a useful benchmark from which to start creating realistic but fake data. We hope that this will provide an even better classifier to detect malware running on IoT devices.

\indent The anomaly detection code is available on GitHub. The README file in the repository has instructions on how to run the code used to generate the graphs and results discussed in this paper.

\printbibliography


\end{document}